\documentstyle[aps,twocolumn,floats,epsf]{revtex}

\begin{document}  
\draft
\title{Gate-Voltage Studies of Discrete Electronic 
States in Al Nanoparticles}                      
         
\author{D.~C.~Ralph\cite{DRnewadd}, 
C.~T.~Black\cite{CBnewadd}, and M. Tinkham}  
\address{Department of Physics
and Division of Engineering and Applied Science, 
Harvard University, 
Cambridge, MA 02138}
\maketitle
     
\begin{abstract}
We have investigated the spectrum of discrete 
electronic states in single, nm-scale Al particles 
incorporated into new tunneling {\em transistors}, 
complete with a gate electrode.  The addition 
of the gate has allowed (a) measurements of 
the electronic spectra for different numbers of 
electrons in the {\em same} particle, (b) greatly 
improved resolution and qualitatively new 
results for spectra within superconducting 
particles, and (c) detailed studies of the 
gate-voltage dependence of the resonance level 
widths, which have directly demonstrated the 
effects of non-equilibrium excitations.
\end{abstract}     
\par                                         
\vspace{0.5cm}


Recently it has become possible to 
measure the discrete spectrum of quantum 
energy levels for the interacting electrons 
within single semiconductor quantum dots \cite{Kastner} 
and nm-scale metal particles 
\cite{Ralph,Black,Agam}, and 
thereby to investigate the forces governing 
electronic structure.  Our earlier experiments 
on Al particles were performed with simple 
tunneling devices, lacking a gate with which 
the electric potential of the particle could be 
adjusted.  In this Letter, we describe the 
fabrication and study of nanoparticle 
{\em transistors}, complete with a gate electrode.  This 
greatly expands the accessible physics.  We 
have used the gate to tune the number of 
electrons in the particle, so as to measure 
excitation spectra for different numbers of 
electrons in the same grain and to confirm 
even-odd effects.  The gate has also allowed 
significantly improved spectroscopic 
resolution, providing new understanding 
about the destruction of superconductivity in a 
nm-scale metal particle by an applied magnetic 
field.  Studies of the gate-voltage dependence 
of tunneling resonance widths have shown that 
non-equilibrium excitations in the nanoparticle 
are a primary source of resonance broadening. 

A schematic cross-section of our device 
geometry is shown in Fig.~1(a).  
The gate 
electrode forms a ring around the Al 
nanoparticle.  The devices are fabricated by 
first using electron beam lithography and 
reactive ion etching to make a bowl-shaped 
hole in a suspended silicon nitride membrane, 
with an orifice between 5 and 10 nm in 
diameter \cite{Ralls}.  The gate electrode is formed by 
evaporating 12 nm of Al onto the flat (lower in 
Fig.~1(a)) side of the membrane.  Plasma 
anodization and deposition of insulating SiO 
are then used to provide electrical isolation for 
the gate.  We next form an aluminum electrode 
which fills the bowl-shaped side (top in 
Fig.~1(a)) of the nitride membrane by evaporation 
of 100 nm of Al, followed by oxidation in 50 
mtorr O$_2$ for 45 sec to form a tunnel barrier 
near the lower opening of the bowl-shaped 
hole.  We create a layer of nanoparticles by 
depositing 2.5 nm of Al onto the lower side of 
the device; due to surface tension the metal 
beads up into separate grains less than 10 nm 
in diameter \cite{Zeller}.  In approximately 25\% 
\begin{figure}
\vspace{0in}
\begin{center}
\leavevmode
    \epsfxsize=2in
 \epsfbox{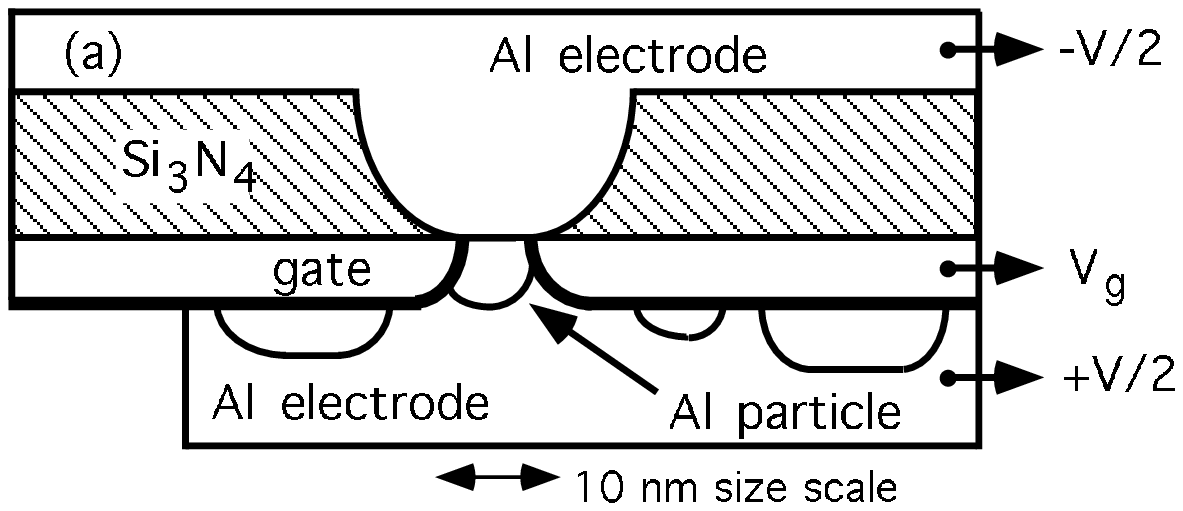} 
 \end{center}
\vspace{-0.39in}
 \begin{center}
 \leavevmode
    \epsfxsize=2.3in
 \epsfbox{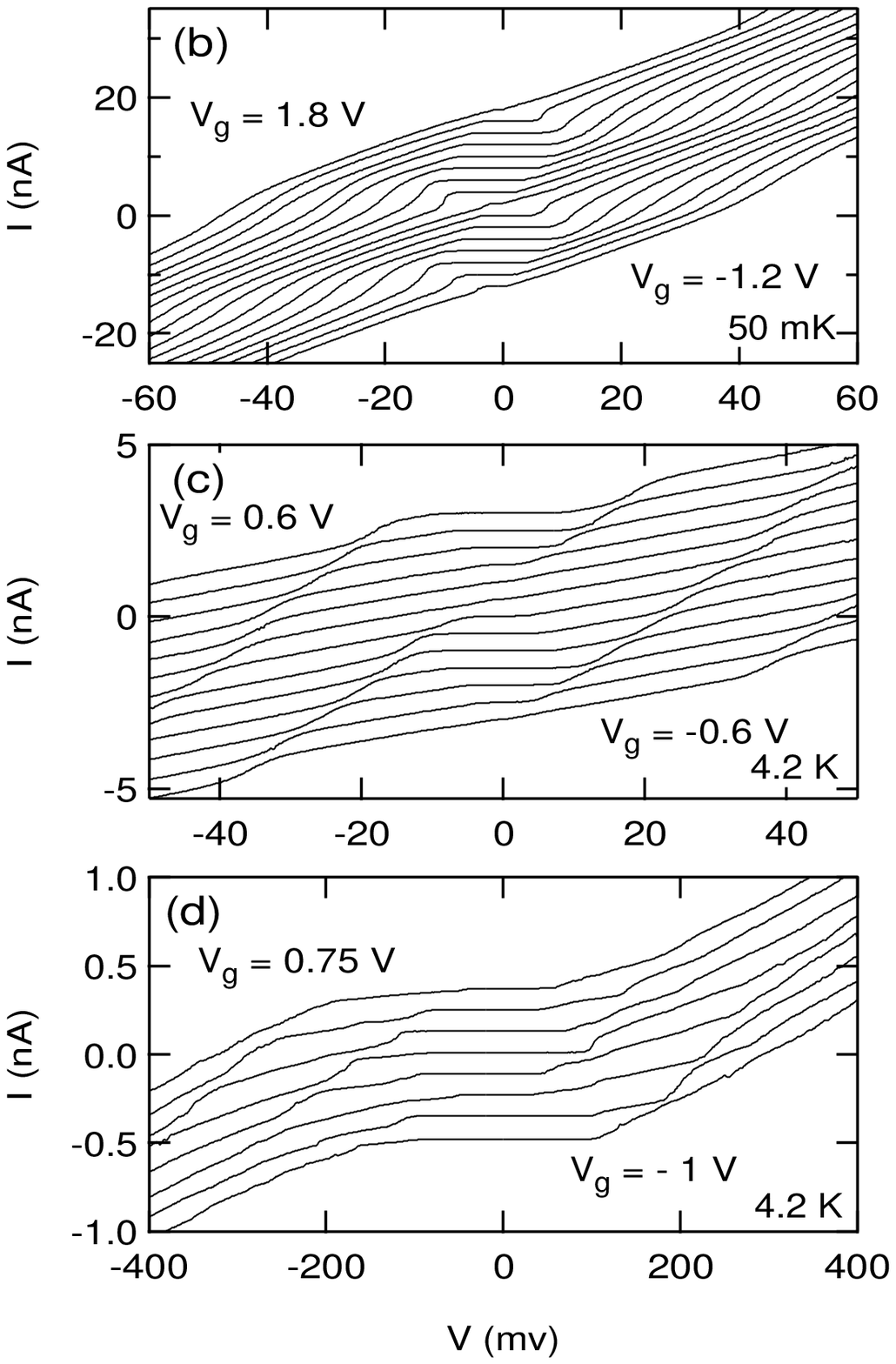}
\end{center}
 \caption{(a) Schematic cross section of device 
geometry.  (b)-(d) Current-voltage curves 
displaying Coulomb-staircase structure for 
three different samples, at equally spaced 
values of gate voltage.  Data for different $V_g$  
are artificially offset on the current axis.}
\end{figure}
of the samples (determined as those showing 
``Coulomb-staircase'' structure as described 
below), a single particle forms under the 
nm-scale tunnel junction to contact the top Al 
electrode.  Finally, after a second oxidation step 
to form a tunnel junction on the exposed 
surface of the particle, a lower electrode is 
formed by evaporating 100 nm of Al to cover 
the particle.  We measure electron tunneling 
between the top and bottom electrodes, 
through a single nanoparticle, as a function of 
gate voltage, $V_g$.

The devices can be characterized by 
measuring large-scale current vs.~source-drain 
voltage ($I$-$V$) curves for a series of $V_g$ 
(Fig.~1(b)-(d)).  The form of these curves, with zero 
current at low  $|V|$  (the ``Coulomb blockade''), 
sloping steps equally spaced in $V$, and step 
thresholds sensitive to $V_g$, is indicative of 
single-electron tunneling via one nanoparticle 
\cite{Grabert}.  From the positions of the voltage 
thresholds for steps in the $I$-$V$ curve, we can 
determine the capacitances within the device 
\cite{Hanna}.  For Fig.~1(b), the lead-to-particle 
capacitances are $C_1$ = 3.5 aF and $C_2$ = 9.4 aF, 
and the gate-to-particle capacitance is $C_g$ = 0.09 
aF;  for Fig.~1(c) the capacitances are 3.4, 8.5, 
and 0.23 aF, and for Fig.~1(d) 0.6, 1.0, and 0.13 
aF.  The charging energy, $E_c = e^2/(2C_{total})$, for 
these devices is relatively large -- for the device 
of Fig.~1(d), $E_c$ = 46 meV (corresponding to T $\sim$ 
500 K), comparable to the largest blockade 
energy measured for any single-electron 
transistor \cite{Takahashi}.

The nanoparticle size can be estimated 
using a value for the capacitance per unit area, 
0.075 aF/nm$^2$, determined from larger tunnel 
junctions made using our oxidation process.  If 
we make the crude assumption of a 
hemispherical particle shape, and base the 
estimate on the larger lead-to-particle 
capacitance, we estimate radii of 4.5, 4.3, and 
1.5 nm, respectively, for Fig.~1(b)-(d).

To measure the discrete electronic states 
within the nanoparticle, we cool the devices to 
mK temperatures and measure $dI/dV$ vs.~$V$ in 
the range of the first Coulomb-staircase step 
(Fig.~2(a)).  
\begin{figure}
  \begin{center}
 \leavevmode
    \epsfxsize=2.5in 
\epsfbox{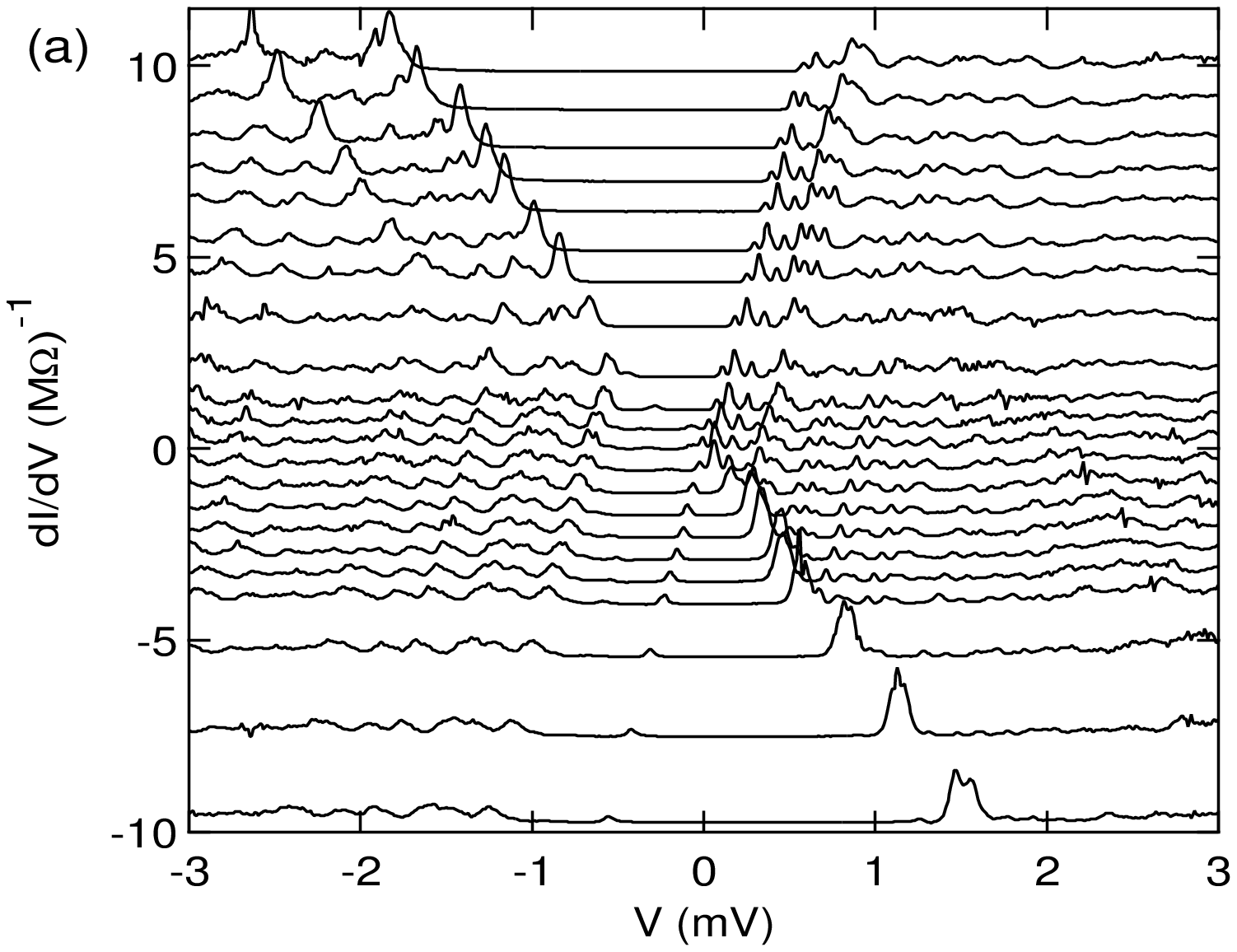}
 \end{center}
\vspace{-0.38in}
  \begin{center}
 \leavevmode
    \epsfxsize=2.5in 
\epsfbox{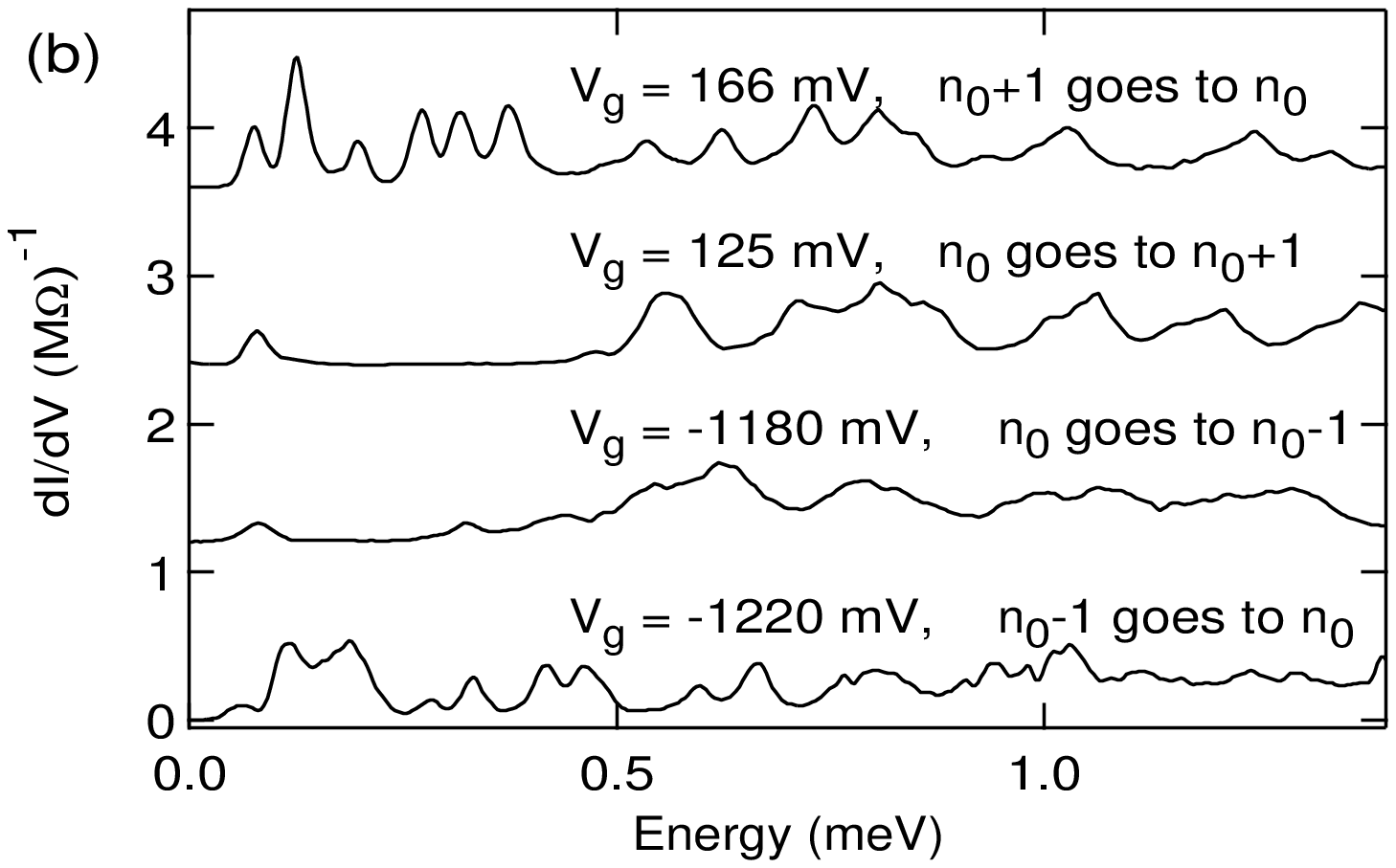}
 \end{center}
\caption{(a) $dI/dV$ vs.~source-drain voltage, 
plotted for $V_g$ ranging from 75 mV (bottom) to 
205 mV (top), for the device of Fig.~1(b).  
Curves are offset on the $dI/dV$ axis.  (b) 
Tunneling spectra for this sample, labeled by 
the number of electrons in the initial and final 
states.  ($n_0$  is odd)  All data are for $T$=50 mK, 
$H$=0.05 Tesla, to drive the Al leads normal.}
\end{figure}
Peaks in $dI/dV$ are expected 
whenever the Fermi level in one of the two 
leads becomes equal to the threshold energy 
for an electron to tunnel either into or out of 
one of the discrete states within the particle, 
through one of the two tunnel junctions \cite{Averin}.  
The interpretation that the data in Fig.~2(a) are 
due to tunneling via states in a single 
nanoparticle is confirmed by the uniform 
shifting of the peaks with $V_g$.  All the $dI/dV$ 
peaks shift linearly in $V$ as a function of $V_g$, 
with tunneling thresholds across junction 1 all 
moving with a single slope, 
$\Delta$$V \approx (C_g/C_2) \Delta$$V_g$, 
and thresholds across junction 2 moving as 
$\Delta$$V \approx (C_g/C_1) \Delta$$V_g$.  
Due to the large charging 
energy in this device, electrons must tunnel 
one at a time through the particle in the $V$ 
range displayed in Fig.~2(a).  This means that at 
fixed $V_g$ all the peaks associated with the same 
junction are due to states with the {\em same} number 
of electrons.

As $V_g$ is increased in Fig.~2(a), the extent 
of the Coulomb blockade region at low $|V|$
decreases, goes to 0, and then increases.  This 
zero-crossing indicates that an electron is 
added to the particle.  If $n_0$  is the number of 
electrons in the ground state at $V_g$=$V$=0, then 
the tunneling processes which overcome the 
Coulomb blockade correspond in the bottom 
half of Fig.~2(a) to $n_0  \rightarrow (n_0 +1)$-electron 
transitions, and in the top half of the figure to 
$(n_0 +1) \rightarrow n_0$  transitions.  
The $n_0  \rightarrow (n_0 +1)$ and 
$(n_0 +1) \rightarrow n_0$-electron spectra can be determined 
most easily from the lower left and upper right 
quadrants of Fig.~2(a), respectively.  In these 
quadrants, the tunneling step which overcomes 
the Coulomb blockade occurs across the higher 
resistance junction \cite{resistance}, so that tunneling 
across this junction is always the rate-limiting 
step for current flow.  All the peaks in $dI/dV$ 
correspond to states which provide alternative 
tunneling channels across this one junction.  In 
the other two quadrants, thresholds for 
tunneling across both tunnel junctions, {\em i.e.~}for 
both $n_0 \rightarrow(n_0 +1)$ and 
$(n_0 +1)\rightarrow n_0$  processes, are 
visible.

In Fig.~2(b), we display several 
tunneling spectra for different numbers of 
electrons in the same particle.  (We discuss the 
significance of these spectra below.)  Fig.~3 
shows the magnetic-field ($H$) dependence of 
the levels which can be resolved in the upper 
two curves of Fig.~2(b).  
\begin{figure}
  \begin{center}
 \leavevmode
    \epsfxsize=2.2in 
\epsfbox{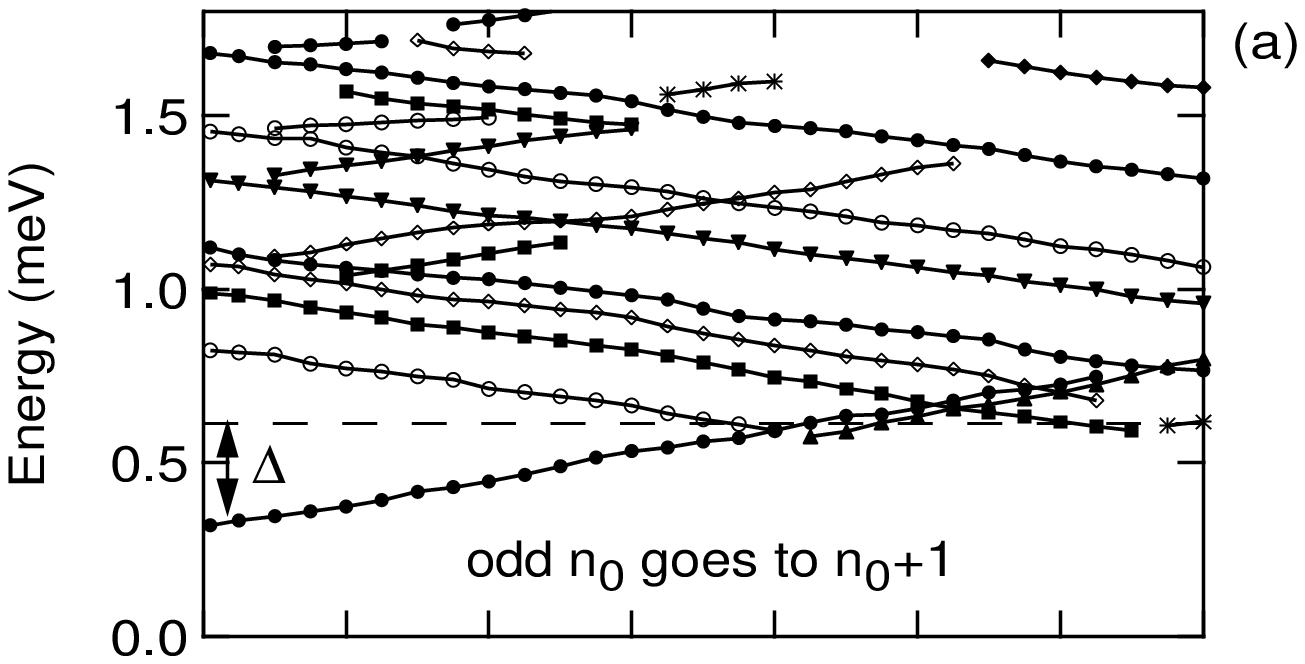}
 \end{center}
\vspace{-0.6in}
  \begin{center}
 \leavevmode
    \epsfxsize=2.2in 
\epsfbox{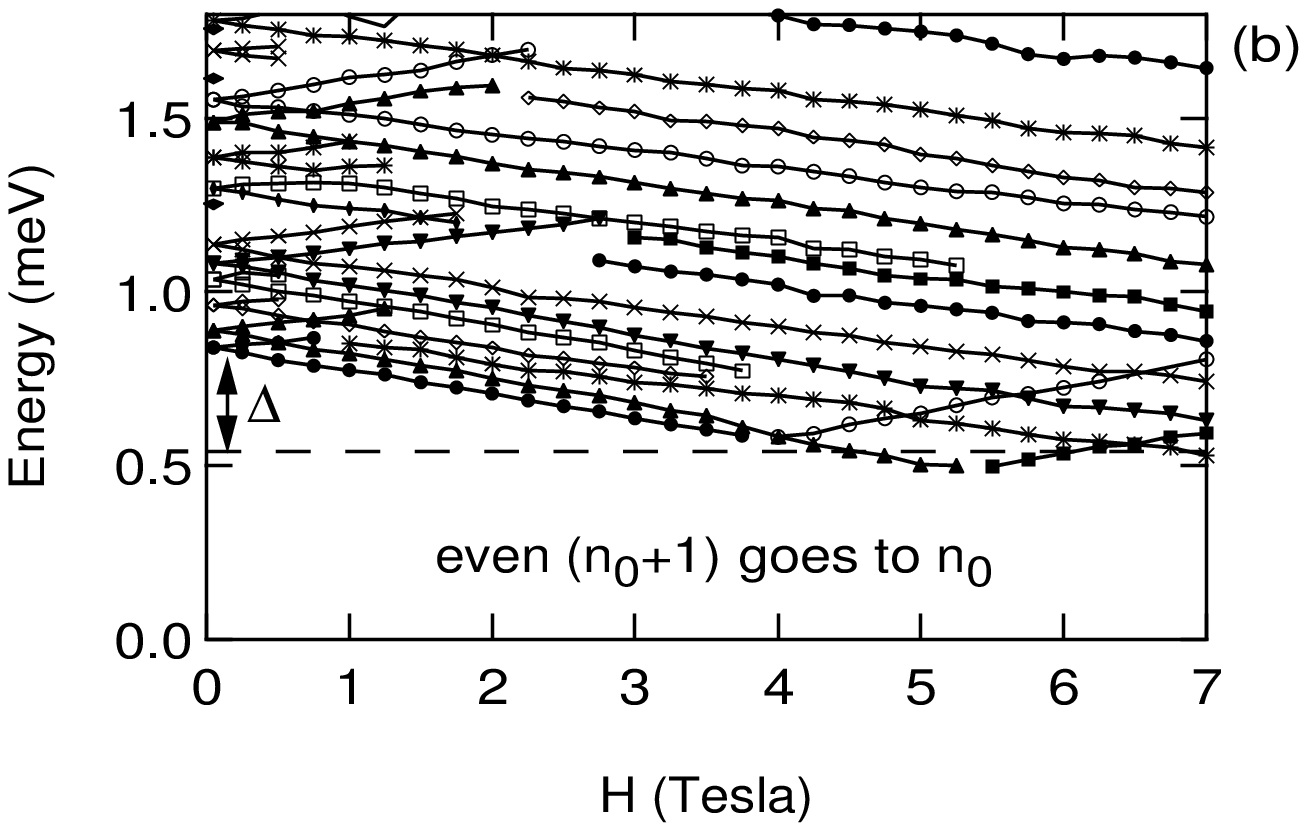}
 \end{center}
\caption{Magnetic field dependence of the 
resolved electronic states for the device of 
Fig.~2 at (a) $V_g$ = 110 mV and (b) $V_g$ = 181 mV.}
\end{figure}
In both Figs.~2(b) and 3, 
we have converted from source-drain bias to 
electron energy, multiplying by $eC_2/(C_1+C_2) = 
0.73e$ to account for the capacitive division of $V$ 
across the two tunnel junctions.  The values of 
$V_g$ were chosen so that the spectra were 
measurable at low values of $|V|$, where they are 
best resolved.  As a function of $H$, the energy 
levels are shifted approximately linearly, with 
Zeeman splittings corresponding to g values 
between 1.95 and 2.  Energy levels which move 
to higher energies with increasing $H$ produce 
broader, less distinct $dI/dV$ peaks than 
downward-moving levels (for reasons poorly 
understood), and can be followed for only a 
limited range of $H$ before they are lost in 
background.  In our previous superconducting 
particles, without gates, most upward-moving 
levels could not be resolved at all.  This caused 
us to incorrectly assign anomalously low 
g-factors to some states \cite{Black}.

As described previously \cite{Ralph}, the 
electron-number parity in the ground state of a 
particle can be determined by whether the 
lowest-energy tunneling level exhibits Zeeman 
splitting (even parity) or not (odd).  In this way 
we can tell that $n_0$  is odd in Fig.~3(a) and $n_0 +1$ 
is even in Fig.~3(b).  While tuning $V_g$, we have 
observed parity changes only when electrons 
are added to the particle at the zero-crossings 
of the Coulomb blockade.  The parity simply 
switches from even to odd to even, etc., at 
consecutive blockade minima.  The 
nanoparticle therefore exchanges electrons 
only with the electrodes, and not with any 
nearby defects.

The ability to tune the number of 
electrons in a particle using $V_g$ allows us for the 
first time to study the dramatic differences 
between tunneling spectra for even and odd 
numbers of electrons in the {\em same} 
superconducting particle \cite{Black}.  The large gap 
between the lowest energy level and all the 
others in the $n_0 \rightarrow n_0 +1$ and 
$n_0 \rightarrow n_0 -1$ spectra 
(Figs.~3(a) and 2(b)) can be explained by 
superconducting pairing.  The tunneling states 
in these spectra have an even number of 
electrons, so that the lowest level is the fully 
paired superconducting state.  Tunneling via 
any other state requires the production of at 
least two quasiparticles, with a large extra 
energy cost approximately twice the 
superconducting gap, $\Delta$.  
The $n_0 +1 \rightarrow n_0$  and 
$n_0 -1 \rightarrow n_0$  tunneling states have an odd number 
of  electrons, and they all must contain at least 
one quasiparticle.  Hence there is no large gap 
within these spectra.  However, the 
contribution of $\Delta$ to the quasiparticle energy 
causes the low-lying tunneling levels at low $H$ 
to have energies greater by $\Delta$ than at large $H$, 
where superconductivity is suppressed.  The 
first tunneling thresholds in Fig.~3(a) and (b) 
have exactly the same $H$ dependence, with 
opposite sign, because they correspond to the 
filling and emptying of the same quantum 
states.  In Fig.~3, $\Delta \approx$  0.3 meV, comparable to 
previous results \cite{Black}.

The levels in Figs.~3(a,b) provide new 
insights as to how a magnetic field destroys 
superconductivity in a nanoparticle.  Consider 
the second level at small $H$ in Fig.~3(a), which 
begins near 0.8 meV.  This state shifts down as 
a function of $H$, due to its spin-1/2 Zeeman 
energy, up to $\sim$ 4 Tesla.  There it disappears in 
favor of a new upward-moving (opposite 
spin-1/2) level.  This means that the 
originally-empty downward-trending level drops below 
the energy of an originally-filled 
upward-trending level, and an electron is transferred 
between the states.  The odd-electron ground 
state changes its spin from 1/2 to 3/2 $\hbar$.  As 
this process is repeated, the tunneling 
threshold moves in a continuous zig-zag 
pattern, and the ground state successively 
increases its spin in units of $\hbar$.  A similar 
argument for Fig.~3(b) shows that the 
even-electron ground state also evolves by 
individual spin flips.  Superconductivity is 
destroyed as electrons flip one at a time.  In 
contrast, the classic theories of Clogston and 
Chandrasekhar \cite{Clogston}, for a superconducting 
transition driven by spin pair-breaking, predict 
a large discontinuous jump in the tunneling 
threshold, at a field where many spins flip 
simultaneously.  Investigations are underway 
as to whether these theories do not properly 
take into account the effect of discrete 
electronic energy levels in the particle \cite{Braun}, or 
perhaps whether the experimental transitions 
are made continuous by a contribution from 
orbital pair breaking \cite{Meservey}.

We have shown that changes in $V_g$ act to 
shift the electrostatic energy of the eigenstates 
on the nanoparticle, and thereby shift the value 
of $V$ at which a given state produces a peak in 
$dI/dV$ (see Fig.~2(a)).  In Fig.~4, we examine 
more closely how the {\em shapes} of resonances 
change as they are moved to larger values of 
$|V|$.  
\begin{figure}
  \begin{center}
 \leavevmode
    \epsfxsize=2.5in 
\epsfbox{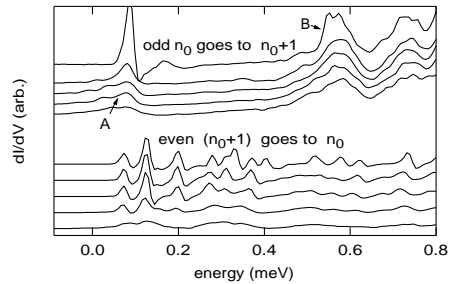}
 \end{center}
\caption{Tunneling resonances broaden and 
develop substructure as $V_g$ is used to shift 
them to larger values of $|V|$.  The top scan in 
both sets corresponds to a value of $V_g$ which 
allows the spectrum to be measured with 
minimum $|V|$.  Going down, other scans are for 
$V_g$ requiring extra source-drain voltages $\delta|V|=$
0.2, 0.4, 0.6, and 1.2 mV.  The scans are 
artificially shifted in energy to align peaks due 
to the same eigenstates.  $T$=50 mK, $H$=0 for all 
scans, and the sample is the same as in Figs.~2 
and 3.  Peaks A and B are discussed in the text.}
\end{figure}
The top spectrum in each set of traces 
corresponds to a value of $V_g$ which places the 
first tunneling peak at the smallest possible $|V|$
\cite{scgap}.  The lower curves show the results of 
tunneling via the same quantum states, after 
their $dI/dV$ peaks have been shifted to 
successively larger $|V|$.  To aid comparison, we 
have aligned the spectra, so that the shifts in 
$|V|$ are not displayed explicitly.  We show data 
for superconducting electrodes ($H$=0) because 
the BCS singularity in the density of states in 
the electrodes improves spectroscopic 
resolution \cite{Ralph}.  Focus first on the lowest energy 
level for $n_0  \rightarrow n_0 +1$ (odd-to-even) electron 
tunneling.  In equilibrium, this is necessarily a 
single, non-degenerate state -- the fully-paired 
superconducting state.  Indeed, for the 
lowest-$|V|$ trace (top curve), this level produces a 
single sharp signal (whose shape is given by 
the derivative of the BCS density of states in 
the Al electrode \cite{Ralph}).  However, as $V_g$ is used to 
shift this first state to larger $|V|$, the resonance 
quickly broadens and develops substructure.  
The substructure cannot be explained by 
heating in the electrodes.  Instead, it is 
evidence for the importance of 
non-equilibrium excitations {\em within} 
the particle \cite{Agam}.  
When electrons are tunneling via even the 
lowest level in the $n_0  \rightarrow n_0 +1$ spectrum, the 
current flow will naturally generate 
non-equilibrium excitations within the particle 
when the excess source-drain energy, $e(\delta|V|)$, 
is greater than the difference between the first 
two levels in the measured $n_0 +1\rightarrow n_0$  spectrum 
($\sim$~0.05 meV).  This happens because one 
electron can tunnel from the high-energy 
electrode into the empty level in the particle, 
and then an electron can exit to the other 
electrode from a {\em different}, lower-energy 
$(n_0 +1)$-electron filled state, leaving a hole.  If the 
resultant electron-hole excitation does not relax 
before the next electron tunnels onto the 
particle ($e/I \sim 10^{-8}$ s), its presence can produce 
small shifts in the tunneling energies available 
for the next electron.  For very small particles, 
it has been proposed that the time-integrated 
result for an ensemble of possible excitations 
will be a well-resolved cluster of tunneling 
resonances associated with each single-electron 
level \cite{Agam}.  For particles large enough to exhibit 
superconductivity, we propose that the 
different non-equilibrium resonances are not so 
well resolved, but overlap to produce 
broadened $dI/dV$ peaks.  

In sharp contrast to the $n_0  \rightarrow n_0 +1$ 
spectrum, the low-energy peaks in the $n_0 +1 \rightarrow 
n_0$  spectrum do not show increasing widths as 
long as  $\delta|V| \leq$ 0.6 meV, but then they do 
broaden for larger $\delta|V|$.  This is additional 
evidence for the role of non-equilibrium 
excitations in resonance broadening, because 
the condition necessary for producing 
non-equilibrium excitations in the particle during 
measurement of these $n_0 +1 \rightarrow n_0$  states is that 
$e(\delta|V|)$ must be greater than the difference 
between the first two levels in the $n_0 \rightarrow n_0 +1$ 
spectrum ($\sim$~0.55 meV).

In the raw data, the first peak (A) in the 
$\delta|V| =$ 0.6 mV scan of the 
$n_0  \rightarrow n_0 +1$ spectrum in 
Fig.~4 and the second peak (B) in the $\delta|V| =$ 0 
scan of the same spectrum occur at the same 
value of $V$, so the degree of non-equilibrium in 
the particle should be similar.  The effective 
widths of these $dI/dV$ peaks are nearly the 
same.  This suggests that the non-equilibrium 
effect is also a dominant source of broadening 
for the higher-energy resonances, even in the 
$\delta|V| =$ 0 spectrum.

In summary, we have produced 
tunneling transistors containing single Al 
nanoparticles, and have measured the discrete 
spectra of energy levels in the particle while 
tuning the number of electrons it contains.  We 
have directly demonstrated differences in level 
spectra for even vs.~odd numbers of electrons, 
which can be explained as an effect of 
superconducting pairing interactions.  The 
application of a magnetic field destroys 
superconductivity in the nanoparticle by a 
sequence of individual spin flips.  The 
tunneling resonances broaden and develop 
substructure when the source-drain voltage    
becomes large enough to allow the production 
of non-equilibrium excitations within the 
particle.

We thank F. Braun and J. von Delft for 
discussions, and R. C. Tiberio for help in device 
fabrication.  This research was supported by 
NSF Grant No. DMR-92-07956, ONR Grant No. 
N00014-96-1-0108, JSEP Grant No. 
N00014-89-J-1023, and ONR AASERT Grant No. 
N00014-94-1-0808, and was performed in part at the 
Cornell Nanofabrication Facility, funded by the 
NSF (Grant No. ECS-9319005), Cornell 
University, and industrial affiliates.


\begin{thebibliography}{99}
\bibitem{resistance} From fits to Fig.~1(b), the 3.5 aF tunnel junction 
has a resistance of 
approximately 2 M$\Omega$, and the 9.4 aF junction 200 k$\Omega$.
\bibitem{scgap} With superconducting electrodes, $e|V| \geq$
twice the gap in the electrodes is 
necessary to initiate tunneling.
\end{thebibliography}
\end{document}